\documentclass[11pt]{article}
\begin{document}
%

%
\def\tilde{\widetilde}
\def\bar{\overline}
\def\hat{\widehat}
\def\*{\star}
\def\[{\left[}
\def\]{\right]}
\def\({\left(}
\def\){\right)}
\def\zb{{\bar{z} }}
\def\frac#1#2{{#1 \over #2}}
\def\inv#1{{1 \over #1}}
\def\half{{1 \over 2}}
\def\d{\partial}
\def\der#1{{\partial \over \partial #1}}
\def\dd#1#2{{\partial #1 \over \partial #2}}
\def\vev#1{\langle #1 \rangle}
\def\bra#1{{\langle #1 |  }}
\def\ket#1{ | #1 \rangle}
\def\rvac{\hbox{$\vert 0\rangle$}}
\def\lvac{\hbox{$\langle 0 \vert $}}
\def\2pi{\hbox{$2\pi i$}}
\def\e#1{{\rm e}^{^{\textstyle #1}}}
\def\grad#1{\,\nabla\!_{{#1}}\,}
\def\dsl{\raise.15ex\hbox{/}\kern-.57em\partial}
\def\Dsl{\,\raise.15ex\hbox{/}\mkern-.13.5mu D}
\def\comm#1#2{ \BBL\ #1\ ,\ #2 \BBR }
\def\x{\stackrel{\otimes}{,}}
\def\det{ {\rm det}}
\def\tr{{\rm tr}}
\def\bacs{\backslash}
\def\tder{\partial_t}
\def\xder{\partial_x}

%
\def\th{\theta}  \def\Th{\Theta}
\def\ga{\gamma}  \def\Ga{\Gamma}
\def\be{\beta}
\def\al{\alpha}
\def\ep{\epsilon}
\def\la{\lambda} \def\La{\Lambda}
\def\de{\delta}  \def\De{\Delta}
\def\om{\omega}  \def\Om{\Omega}
\def\sig{\sigma} \def\Sig{\Sigma}
\def\vphi{\varphi}
%
%
\def\CA{{\cal A}} \def\CB{{\cal B}} \def\CC{{\cal C}}
\def\CD{{\cal D}} \def\CE{{\cal E}} \def\CF{{\cal F}}
\def\CG{{\cal G}} \def\CH{{\cal H}} \def\CI{{\cal J}}
\def\CJ{{\cal J}} \def\CK{{\cal K}} \def\CL{{\cal L}}
\def\CM{{\cal M}} \def\CN{{\cal N}} \def\CO{{\cal O}}
\def\CP{{\cal P}} \def\CQ{{\cal Q}} \def\CR{{\cal R}}
\def\CS{{\cal S}} \def\CT{{\cal T}} \def\CU{{\cal U}}
\def\CV{{\cal V}} \def\CW{{\cal W}} \def\CX{{\cal X}}
\def\CY{{\cal Y}} \def\CZ{{\cal Z}}
%
%
\catcode`\@=11
\font\tenmsa=msam10 at 12truept
\font\sevenmsa=msam7
\font\fivemsa=msam5
\font\tenmsb=msbm10 at 12truept
\font\sevenmsb=msbm7 at 9truept
\font\fivemsb=msbm5 at 7truept
\newfam\msafam
\newfam\msbfam
\textfont\msafam=\tenmsa \scriptfont\msafam=\sevenmsa
 \scriptscriptfont\msafam=\fivemsa
\textfont\msbfam=\tenmsb \scriptfont\msbfam=\sevenmsb
 \scriptscriptfont\msbfam=\fivemsb
 
\def\hexnumber@#1{\ifcase#1 0\or1\or2\or3\or4\or5\or6\or7\or8\or9\or
        A\or B\or C\or D\or E\or F\fi }
 
\def\msb{\tenmsb\fam\msbfam}
\def\Bbb{\ifmmode\let\next\Bbb@\else
 \def\next{\errmessage{Use \string\Bbb\space only in math mode}}\fi\next}
\def\Bbb@#1{{\Bbb@@{#1}}}
\def\Bbb@@#1{\fam\msbfam#1}
\def\Cmath{\Bbb C}
\def\Rmath{\Bbb R}
\def\Zmath{\Bbb Z}
\def\cadremath#1{\vbox{\hrule\hbox{\vrule\kern8pt\vbox{\kern8pt
   \hbox{$\displaystyle #1$}\kern8pt} 
   \kern8pt\vrule}\hrule}}
\def\proof{\noindent Proof. \hfill \break}
\def\cqfd{ {\hfill{$\Box$}} }
\def\square{\hfill
\vrule height6pt width6pt depth1pt \\}
\def\refp#1{(\ref{#1})}
%
%
\def\debut{ \begin{eqnarray} }
\def\fin{ \end{eqnarray} }
\def\non{ \nonumber }
%

%
%
%
\def\bacs{\backslash}

\rightline{T00/024}
\vskip 1cm
\centerline{\LARGE Poisson Algebra of}
\bigskip
\centerline{\LARGE  2d Dimensionally Reduced Gravity.}
\vskip 1cm

\vskip1cm

\centerline{\large  Denis Bernard
\footnote[1]{Member of the CNRS; dbernard@spht.saclay.cea.fr}
        and Nicolas Regnault
\footnote[2]{regnault@spht.saclay.cea.fr} }
\centerline{Service de Physique Th\'eorique de Saclay}
\centerline{F-91191, Gif-sur-Yvette, France.}

\vskip2cm

Abstract.\\
Using a Lax pair based on twisted affine $sl(2,R)$ Kac-Moody and Virasoro algebras, we deduce a $r-$matrix formulation 
of two dimensional reduced vacuum Einstein's equations. Whereas the fundamental Poisson brackets are non-ultralocal, 
they lead to pure c-number modified Yang-Baxter equations. We also describe how to obtain classical observables by 
imposing reasonable boundary conditions.

\vfill

\newpage

\section{Introduction}
\label{Intro}

Quantization of gravitation is still today, one of the most challenging problem in theoretical physic. A
 possible approach to have an idea of what happens when passing from classical theory to the quantum one
consists in trying simpler models. The case of two commuting Killing vector reduction of source-free 
general relativity is probably one of the most interesting. This model has the particularity to exhibit 
an infinite dimensional symmetry group, the so-called Geroch group \cite{Geroch}. It is known to be 
integrable since the works of Belinskii, Zakharov and Maison \cite{BelZak, Maison}. One way to quantize 
is to apply methods used for standard integrable models. This quantization problem, which is equivalent
 to ${\bf SL(2,R)/SO(2)}$ coset space $\sigma$-models coupled to two-dimensional gravity and a dilaton 
(it can be generalized to ${\bf G/H}$ coset space), has been of course the subject of several papers 
(\cite{NicoKoroSamt1, KoroSamt1} and references therein).

Before trying to quantize this system, it is necessary to study in details the Poisson algebra. 
The classical theory is also interesting for its own. An extensive work has already 
been done in this domain by Julia, Korotkin, Nicolai and Samtleben \cite{NicoKoroSamt1, KoroSamt1, 
JulNico}. A dynamical $r-$matrix formulation of the model has been proposed (see e.g. \cite{KoroSamt1})
 where the dilaton is given a priori. A restriction of this approach is that the brackets are 
evaluated on the constraint surface, which prevents deduction of the associated Yang-Baxter 
equations. In \cite{BerJul} an expression was proposed for the Lax connection based on twisted
 $sl\left(2, R\right)$ affine Kac-Moody and Virasoro algebras that reproduces the equations of 
motion. Using dressing transformations, it provides a rather elegant method to generate solutions
 \cite{BerReg}. The aim of this article is to show that this form of the Lax connection can also
 provide a good basis to obtain a $r-$matrix formulation of this problem. This means that all fields
 are considered as dynamical variables and pure $c$-number Yang-Baxter equations can be deduced. 
The structure we obtain is closed to Toda affine model's one. We hope that we will be able to 
transpose what have been done in this domain to our problem. This will provide an alternative 
algebraic approach for the quantization of 2d reduced gravity and thus a complementary point 
of view to what has already been done.

This paper is organized as follow. Section \ref{LaxEq} sums up some of the main results of 
\cite{BerJul}, in particular we introduce the Lax pair. Section \ref{Poisson} deals with the 
Hamiltonian formulation of the theory. The calculation of the Poisson brackets of the Lax 
connection, the key point of this paper, and the deduction of the associated Yang-Baxter equations 
are described in section \ref{FPR}. We show that despite the $r-$dynamical behavior of the model, we
 obtain pure $c$-number modified Yang-Baxter relations of kind 
\debut
[r_{12}^{\epsilon_1},r_{23}^{\epsilon_2}]
        +[s_{23}^{\epsilon_2},s_{31}^{\epsilon_3}]
        +[s_{31}^{\epsilon_3},r_{12}^{\epsilon_1}]
        -\frac{1}{2}k_2s^{\epsilon_3}_{31}-\frac{1}{2}k_3r^{\epsilon_1}_{12}
        -\frac{1}{4}\left[U_{23},c_{12}\right]&=&0\non
\fin
that can be interpreted as consistency conditions for a simpler static linear model (but still 
lack of a quadratic interpretation). As an application of the previous results, we determine the
 Poisson brackets of monodromy matrices in section \ref{Mono}, and we point out the problem of 
coincident points. Section \ref{Obs} is an attempt to find classical observables. We show that 
if we impose reasonable boundaries conditions, it is possible to construct an infinite set of 
these objects. Finally, we have gathered in Appendix A and B, some  expressions and sketches of 
demonstrations related to section \ref{FPR}.

\section{Equation of motion and Lax connection}
\label{LaxEq}

     In this section, we will review some basics facts and results found in \cite{BerJul}. We will
 also introduce notations used in this article. Let us recall the parameterization we choose for the metric :
\debut
ds^2=\rho^{\half}e^{2\hat{\sigma}}(-dt^2+dx^2)+\rho S_{ij}(x,t)dy^idy^j
\fin 
where $\rho$ is called the dilaton and $\hat\sigma$ the conformal factor. The symmetric $2 \times 2$
 matrix $S$, normalized by $det(S)=1$, can be written as $S=\CV^t\CV$, where $\CV$ is an element of 
$SL\left(2, R\right)$. $\CV$ is equivalent to internal zweibein, up to a $\sqrt{\rho}$ factor. There 
is a manifest local $SO(2)$ gauge symmetry when multiplying $\CV$ to the left with any element of $SO(2)$.

We introduce the decomposition of $sl(2,R)=h \oplus g$ where $h=so(2)$ is the maximal compact subalgebra 
of $sl(2,R)$. We will use the following notation for the generators : $T^\alpha$ with a Greek index 
correspond to the generator of $h$ and $T^a$ with a Latin index correspond to the generators of $g$.
 We choose these generators such that they are orthogonal and normalized with respect to the Killing form. 
To formulate the vacuum Einstein's equation (the so-called Ernst's equations), we introduce the connection 
$\CV\partial\CV^{-1}$. We denote each component of this connection as $P_x+Q_x=\CV\xder\CV^{-1}$ and 
$P_t+Q_t=\CV\tder\CV^{-1}$, where $P$ is an element of $g$ and $Q$ belong to $h$. With these objects, 
Einstein's equation can be brought to the form

\debut
\xder Q_t - \tder Q_x +\left[Q_x,Q_t\right]+\left[P_x,P_t\right]&=&0\label{motion1}\\
\xder P_t + \left[Q_x,P_t\right] &=& \tder P_x + \left[Q_t,P_x\right]\label{motion2}\\
\xder \left(\rho P_x\right) + \left[Q_x,\rho P_x\right] &=& \tder \left(\rho P_t\right) 
+ \left[Q_t,\rho P_t\right]\label{motion3}\\
\left(\tder^2 - \xder^2\right)\rho&=&0\label{motion4}\\
\left(\left(\tder \pm \xder\right)\rho\right)\left(\tder \pm \xder\right)\hat{\sigma}
&=&-\rho\half tr\left(\left(P_x \pm P_t\right)^2\right)\label{fosig}
\fin

Before dealing with the Lax connection, we shall introduce the algebra we will use. Consider the 
$sl(2,R)$ affine Kac-Moody algebra defined by the commutation relations : 
\debut
\left[X\lambda^n, Y\lambda^m\right]&\ =&  [X,Y]\lambda^{m+n}+
        n\frac{k}{2}tr\left(XY\right)\delta_{n+m,0}
\fin
We twist this algebra with the order two automorphism that leaves $h$ invariant. It means that for 
some element $X\lambda^n$, if $n$ is even, then $X$ is an element of $h$, else $X$ is an element 
of $g$. In fact, we will use the semi direct product of this algebra with the Virasoro's one. We 
recall that the commutation relations for the Virasoro algebra are
\debut
\left[L_n,L_m\right]&=& (n-m)L_{m+n} +
        n(n^2-1)\frac{c}{12}\delta_{n+m,0}\non
\fin
and the crossed Lie bracket is $\left[L_n,X\lambda^m\right]= -\frac{m}{2} X\lambda^{n+m}$. For convenience, 
we introduce a particular notation for two elements of the Virasoro algebra $E_\pm=L_0-L_{\pm 1}$ which 
verify the commutation relation $[E_+,E_-]=E_++E_-$.

As we said before, the model is integrable. It means that an auxiliary linear system
\debut
\left(\tder+A_t\right)\Psi=0&{\rm{and}}&\left(\xder+A_x\right)\Psi=0\label{lin}
\fin
can be found such that the zero curvature condition $\left[\tder+A_t,\xder+A_x\right]=0$ reproduces the 
equations of motion. The expression of the components of Lax connection that fulfills this requirement, is

\debut 
A_x& =& -\frac{1}{2}\rho^{-1}(\Pi_{\hat\sigma}-\xder \rho)E_+
        -\frac{1}{2}\rho^{-1}(\Pi_{\hat\sigma}+\xder \rho)E_-
        + \frac{1}{2}\left(P_{xa}+P_{ta}\right)T^a\lambda\non\\
        &&+\frac{1}{2}\left(P_{xa}-P_{ta}\right)T^a\lambda^{-1}
        + Q_{x\alpha}T^\alpha+\Pi_{\rho}\frac{k}{2}\label{Ax}
\fin

\debut 
A_t&=& -\frac{1}{2}\rho^{-1}(\Pi_{\hat\sigma}-\xder \rho)E_+
        +\frac{1}{2}\rho^{-1}(\Pi_{\hat\sigma}+\xder \rho)E_-
        + \frac{1}{2}\left(P_{xa}+P_{ta}\right)T^a\lambda\non\\
        &&-\frac{1}{2}\left(P_{xa}-P_{ta}\right)T^a\lambda^{-1}
        + Q_{t\alpha}T^\alpha-\xder\hat{\sigma}\frac{k}{2}\label{At}
\fin
where $\Pi_{\hat\sigma}=-\tder\rho$ and $\Pi_{\rho}=-\tder{\hat\sigma}$.

We will often use the notation $A$ for $A_x$ and the term connection for the Lax connection (from
 now, we will deal no more with the connection $\CV\partial\CV^{-1}$, so there will be no misunderstanding). 
Notice that the zero curvature condition reproduces the equations of motion (\ref{motion1}) to 
(\ref{motion4}) and a second order equation for $\hat{\sigma}$
\debut 
\left(\tder^2 - \xder^2\right)\hat{\sigma}&=&-\half tr\left(P_x^2-P_t^2\right)\label{sosig}
\fin
which is a consequence of the two linear equations for the conformal factor (\ref{fosig}).

This connection was first used as a powerful method to generate solutions to 
Einstein's equations. Readers who are interested, can found details in \cite{BerJul} and \cite{BerReg}. 
When comparing with other Lax connection used in the literature (see e.g.\cite{NicoKoroSamt1, KoroSamt1}),
 we remark that all fields are considered on an equal footing (in general, the dilaton is supposed to 
be given and the conformal factor is deduced from the other variables). So, if we want to consider all
 of them at the same time, this connection seems to be a good candidate. We will see that it introduces 
no additional difficulty and, at the contrary, yields simpler and more compact expressions for the 
Poisson brackets for the connection.

\section{Action and canonical brackets}
\label{Poisson}

The hamiltonian formulation of 2d reduced gravity, has already been studied in various papers 
\cite{JulNico, NicoKoroSamt1, KoroSamt1}. So, up to some minor changes, the formulation we shall 
use is identical to what can be found in the literature. Let us recall it.

First of all, let describes our phases space. It is defined by the canonical variables $P_x,Q_x,\rho,\hat\sigma$ 
and their associated momenta $\Pi_P,\Pi_Q,\Pi_{\rho},\Pi_{\hat\sigma}$. Here we use the canonical Poisson 
brackets to define the symplectic structure ($\left\{\rho(x),\Pi_{\rho}(y)\right\}=\delta (x-y)$ and so on).

To make contact with the model, we have to express the quantities $P_t$ and $Q_t$ in terms of the canonical
 variables. We also need  the generators of the transformations associated to the invariances of our system 
which  are invariance under reparameterization and the local $SO(2)$ invariance. To solve this problem, we 
can either try to deduce these formulae from some mathematical procedures (see \cite{KoroSamt1} for example), 
or just give expressions as definition and verify if this choice is coherent. Here, we will adopt the second 
method. First we define the variable $P_t$ as
\debut
-\rho P_t&=&\xder \Pi_P + \left[Q_x,\Pi_P\right]
         + \left[P_x,\Pi_Q\right]
\fin
$Q_t$ will be considered to have vanishing brackets with all variables.

This phase space is reduced by the three constraints arising from the gauge invariance mentioned above. 
First we have the Hamiltonian ${\cal{H}}$ which can be written as 
\debut
\cal{H}&=&- \Pi_\rho \Pi_{\hat\sigma} - \xder \rho \xder \hat\sigma
        + \frac{1}{2} \rho tr\left(P_t^2+P_x^2\right)
        + tr\left(Q_t \Phi\right)\label{H}
\fin
With this expression for the Hamiltonian, the equations of motion (\ref{motion1}-\ref{motion4} and \ref{sosig}) 
are correctly reproduced (the brackets needed for these calculations are given below). We still have two other 
generators of gauge transformations. We denote ${\cal{P}}$ the generator of diffeomorphisms in the spatial 
direction whose expression is
\debut
\cal{P}&=& \Pi_\rho \xder\rho + \Pi_{\hat\sigma} \xder\hat\sigma
        + \rho tr\left(P_t P_x\right)+ tr\left(Q_x \Phi\right)\label{P}
\fin
The linear combinations ${\cal C_\pm=H \pm P} \approx 0$ of these two constraints are equivalent 
to the two linear equations for the conformal factor (\ref{fosig}). Finally, the generator $\Phi$ of the $SO(2)$ 
gauge invariance takes the following form
\debut
-\Phi&=&\xder Q_x + \left[Q_x,\Pi_Q\right]
         + \left[P_x,\Pi_P\right] \label{phi}
\fin
It could be easily verified that $\Phi$ belongs to $so(2)$. Notice that all these constraints are first class 
constraints.

We will use the standard index-free tensor notation. For some element $X$, we define $X_1 \equiv X\otimes I$ 
and $X_2 \equiv I \otimes X$. We will also introduce the decomposition of the Casimir element ${\cal{C}}_{12}$
 of $sl(2,R)$: ${\cal{C}}_{12}=c_{12}+d_{12}$ with $c_{12} = T^\alpha \otimes T_\alpha$ and $d_{12} = T^a \otimes T_a$. 
The validity of this decomposition is due to orthogonality of generators with respect to the Killing form.

We shall list all the basic Poisson brackets needed for further calculations.
\debut 
\left\{P_{t1}(x),P_{t2}(y)\right\}&=&\rho^{-2}(x)\delta(x-y)
        \left[d_{12},\Phi_2(x)\right]\non\\
\left\{P_{t1}(x),P_{x2}(y)\right\}&=&\rho^{-1}(x)\delta'(x-y)d_{12}+
        \rho^{-1}(x)\delta(x-y)\left[d_{12},Q_{x2}(x)\right]\non\\
\left\{P_{t1}(x),Q_{x2}(y)\right\}&=&
        \rho^{-1}(x)\delta(x-y)\left[d_{12},P_{x2}(x)\right]\non\\
\left\{\Phi_{1}(x),\Phi_{2}(y)\right\}&=&\delta(x-y)
        \left[c_{12},\Phi_2(x)\right]\non\\
\left\{P_{t1}(x),\Phi_{2}(y)\right\}&=&-\delta(x-y)
        \left[c_{12},P_{t1}(x)\right]\non\\
\left\{P_{x1}(x),\Phi_{2}(y)\right\}&=&-\delta(x-y)
        \left[c_{12},P_{x1}(x)\right]\non\\
\left\{Q_{x1}(x),\Phi_{2}(y)\right\}&=&\delta'(x-y)c_{12}+
        \delta(x-y)\left[c_{12},Q_{x2}(x)\right]\non
\fin
The four last commutators show that $\Phi$ is the generator of the local $SO(2)$ invariance. Equivalent calculations 
can be done to prove other expressions.
 
\section{Poisson brackets for the Lax connection}
\label{FPR}

\subsection{Main result}

Now that we have defined the Poisson algebra, we can deduce the Poisson bracket for the Lax connection 
(the so-called fundamental Poisson brackets). The raw formula derived from a direct calculation, is quite 
long and has no pedagogic interest. We won't write its expression (readers who want to obtain it will 
encounter no difficulty). What is really interesting is that it can be put on a $r-$matrix form. A brief 
survey of the method used to find this expression is described in Appendix A. The result we have obtained is

\debut
\left\{A_1(x),A_2(y)\right\}
        &=&\frac{1}{\rho(x)}\;\delta(x-y)\;
        \big(\left[r^\epsilon_{12},A_1(x)\right]\;+\;
        \left[s^\epsilon_{12},A_2(x)\right]\big)\non\\
        &+&\left(\frac{1}{\rho(x)}\;s_{12}\;-\;\frac{1}{\rho(y)}\;r_{12}\right) 
        \;\partial_{x}\delta(x-y) \label{fpb}\\
        &-&\frac{1}{8}\frac{1}{\rho^2(x)}\;\delta(x-y)\;
        \left[U_{12},\Phi_{1}(x)-\Phi_{2}(x)\right]\non
\fin
where
\debut
U_{12}&=&d_{12} (\lambda_1-\lambda_1^{-1})(\lambda_2-\lambda_2^{-1})\non
\fin

This formula is the one obtained in the case of non-ultralocal theories \cite{Maillet1} with an additional 
term coming from the local $SO(2)$ invariance. This term has to be considered with caution when dealing with 
the Jacobi identity. Here are the expressions of the $r-$ and $s-$matrices :
 
\debut
r^\pm_{12}&=&\frac{1}{2}
        \frac{(1-\lambda_1^2)(1-\lambda_2^2)}{\lambda_1^2-\lambda_2^2} c_{12}
        +\frac{1}{2}\frac{\lambda_1\lambda_2^{-1}(1-\lambda_2^2)^2}
        {\lambda_1^2-\lambda_2^2}d_{12}
        \mp\frac{1}{2}\left(E_\pm \otimes k+\frac{1}{2}k \otimes 
        \left(E_++E_-\right)\right) \label{r+-}\\
s^\pm_{12}&=&\frac{1}{2}
        \frac{(1-\lambda_1^2)(1-\lambda_2^2)}{\lambda_1^2-\lambda_2^2} c_{12}
        +\frac{1}{2}\frac{\lambda_1^{-1}\lambda_2(1-\lambda_1^2)^2}{\lambda_1^2-\lambda_2^2}d_{12}
        \mp \frac{1}{2}\left(k \otimes E_\mp+\frac{1}{2} 
        \left(E_++E_-\right) \otimes k\right) \label{s+-}
\fin
The matrices involved here are pure c-number, all coordinates dependencies have been factorized in 
the $\rho^{-1}$ factors. The rational functions that appear in \refp{r+-} and \refp{s+-}, have only
 a meaning as formal power series. So, whether we choose $|\lambda_1|<|\lambda_2|$ or $|\lambda_1|>|\lambda_2|$ 
when developing, we obtain two different sets of matrices (here + convention refers to the case 
$|\lambda_1|<|\lambda_2|$). Fully developed formulas are given in appendix A.

Although our Lax connection is not exactly the same that the one used by Korotkin and Samtleben in 
\cite{KoroSamt1}, it however is possible to compare some pieces of \refp{fpb} with their expression. 
The algebra we used is just a way to eliminate the coordinates dependence of the moving poles. In order 
to compare the two formulae we have to restore this dependence. It can be achieved by formally substituting 
$\frac{1-\gamma}{1+\gamma}$ by $\lambda$ where $\gamma$ is the moving pole (a detailed explanation of 
this equivalence can be found in \cite{BerReg}). One can now see, their ultralocal part of the Poisson 
brackets for the connection is equivalent to ours if we just keep the loop part of the $r-$matrices 
(remember that the introduction of the central extension $k$ and the Virasoro algebra is a method to 
take $\rho$ and $\hat{\sigma}$ into account). The case of the non-ultralocal part is more difficult. 
No direct comparison can be done because the dilaton produces additional terms. Notice that their
 brackets is calculated on the constraint surface, thus they have no additional term involving $\Phi$ and
 they can't explicitly verify the Jacobi identity (which imply that Yang-Baxter equations can't be found).

Now, the fundamental Poisson brackets have to satisfy the standard relations of Poisson brackets and to be 
independent of the convention we choose. We will focus here on antisymmetry and independence, dealing 
with the Jacobi identity in the next subsection. Proofs of these two properties lie on the same relations
 of the $r-$matrices, that can be divided into two sets. On one hand, there is a set a pure numerical identities :
\debut
r_{12}^\epsilon&=&-s_{21}^{-\epsilon}\non\\
r_{12}^\epsilon-r_{12}^{-\epsilon}&=&s_{12}^\epsilon-s_{12}^{-\epsilon}\label{rel1}\\
U_{12}&=&U_{21}\non
\fin
and on the other hand, we have got a relation involving connection and dilaton :
\debut
[r_{12}^\epsilon-r_{12}^{-\epsilon},A_1+A_2]&=&-\left(\rho^{-1}
\xder\right)\rho(r_{12}^\epsilon-r_{12}^{-\epsilon})\label{rel2}
\fin
Using these formulae, antisymmetry and independence can be easily shown. For example, from 
eq.\refp{rel1}, we have
\debut
\left\{A_1(x),A_2(y)\right\}\Big|_{\epsilon=+}-\left\{A_1(x),A_2(y)\right\}\Big|_{\epsilon=-}
&=&\rho^{-1}(x)[r_{12}^+-r_{12}^-,A_1(x)+A_2(x)]\;\delta(x-y)\non\\
	&&+\;\left(r_{12}^+-r_{12}^-\right)\;(\rho^{-1}(x)-\rho^{-1}(y))\;\partial_{x}\delta(x-y)\non
\fin
Now with eq.\refp{rel2} and the identity 
$(\rho^{-1}(x)-\rho^{-1}(y))\;\partial_{x}\delta(x-y)\;=\;\rho^{-2}(x)\;\partial_{x}\rho\;\;\delta(x-y)$,
 we easily prove that the right hand-side of the above equation vanishes.

Notice that in more conventional cases like Toda field theories, only numerical relations are used. 
In particular the difference $r_{12}^+-r_{12}^-$ is generally proportional to the Casimir tensor which 
would simplify \refp{rel2}. This  more complicated form is a consequence of the $\rho^{-1}$ terms in \refp{fpb}.

\subsection{Jacobi identity and Yang-Baxter equations}

Finally, we have to prove the validity of the Jacobi identity. Performing the calculation leads to the following formula :

\debut
\left\{A_1(x),\left\{A_2(y),A_3(z)\right\}\right\}+{\rm{perm.}}
        &=&\rho^{-2}(x)\delta(x-y)\delta(y-z)\left(
        \left[A_1(x),A_{123}\right]+{\rm{perm.}}\right)\non\\
        &+&\left(\xder\rho^{-2}\right)\delta(x-y)\delta(y-z)
        \left(B_{123}+{\rm{perm.}}\right)\label{Jac} \\
        &+&\rho^{-3}(x)\delta(x-y)\delta(y-z)\left(
        \left[\Phi_1(x),C_{123}\right]+{\rm{perm.}}\right)\non\\
        &+&\rho^{-2}(x)\delta(y-z)\xder\delta(x-y)D_{123}+{\rm{perm.}}
        \non    
\fin

Explicit expressions of $A_{123}$, $B_{123}$, $C_{123}$ and $D_{123}$ are gathered in appendix B. The fact
 that eq.\refp{Jac} is equal to zero, is a consequence of the properties of $A_{123}$ and $C_{123}$. One
 can shown they are invariant under cyclic permutations and equal to zero (we left details in appendix B). 
This leads to the modified Yang-Baxter equations

\debut
[r_{12}^{\epsilon_1},r_{23}^{\epsilon_2}]
        +[s_{23}^{\epsilon_2},s_{31}^{\epsilon_3}]
        +[s_{31}^{\epsilon_3},r_{12}^{\epsilon_1}]
        -\frac{1}{2}k_2s^{\epsilon_3}_{31}-\frac{1}{2}k_3r^{\epsilon_1}_{12}
        -\frac{1}{4}\left[U_{23},c_{12}\right]&=&0\label{YB1}\\
\left[r_{23}^{\epsilon_2},U_{12}\right]
        +[s_{23}^{\epsilon_2},U_{13}]
        +\frac{1}{2}k_3U_{12}-\frac{1}{2}k_2U_{13}&=&0\label{YB2}
\fin
Notice that the choice of the three conventions has to be consistent. This condition can be written as 
$\left|\epsilon_1+\epsilon_2+\epsilon_3\right|=1$.\\
Thus, the validity conditions of the Jacobi identity are pure c-numbers equations. It could seem to be a 
miracle, especially when looking at first sight at equation \refp{fpb} with its explicit $\rho$ dependence. 
Actually this dependence is encoded in the terms involving the central extension. The $SO(2)$ gauge 
invariance generates the term $\left[U_{23},c_{12}\right]$ and eq.\refp{YB2}. If we want to compare 
these Yang-Baxter equations with those obtained in simpler cases, we have to drop the central extension
 and the local $SO(2)$ constraint. In this case, we obtain the same results as in the case of non-ultralocal 
theories with constant $r-$matrices (see \cite{Maillet1}). 

A possible interpretation of these Yang-Baxter equations is as consistency conditions for linear Poisson 
brackets involving three objects $L^\pm$ and $\phi$. Defining the following algebra

\debut
\left\{L_1^\pm,L_2^\pm\right\}&=&\left[r^\epsilon_{12},L_1^\pm\right]\;+\;\left[s^\epsilon_{12},L_2^\pm\right]
	\;+\;\frac{1}{2}k_2L_1^\pm\;-\;\frac{1}{2}k_1L_2^\pm
        -\frac{1}{8}\left[U_{12},\phi_1-\phi_2\right]\non\\
\left\{L_1^+,L_2^-\right\}&=&\left[r^-_{12},L_1^+\right]\;+\;\left[s^-_{12},L_2^-\right]
	\;+\;\frac{1}{2}k_2L_1^+\;-\;\frac{1}{2}k_1L_2^-
        \;-\;\frac{1}{8}\left[U_{12},\phi_1-\phi_2\right]\non\\
\left\{L_1^-,L_2^+\right\}&=&\left[r^+_{12},L_1^-\right]\;+\;\left[s^+_{12},L_2^+\right]
	\;+\;\frac{1}{2}k_2L_1^-\;-\;\frac{1}{2}k_1L_2^+
        \;-\;\frac{1}{8}\left[U_{12},\phi_1-\phi_2\right]\non\\
\left\{L_1^\pm,\phi_2\right\}&=&\frac{1}{2}\;\left[L_1^\pm-L_2^\pm,c_{12}\right]\non\\
\left\{\phi_1,\phi_2\right\}&=&\frac{1}{2}\;\left[\phi_1-\phi_2,c_{12}\right]\non
\fin

We suppose that $k$ commutes with all other elements. $U_{12}$ and $c_{12}$ are considered to be 
symmetric when permuting the two spaces. $r_{12}$ and $s_{12}$ have to fulfilled \refp{rel1} and 
\refp{rel2} with $\rho$ constant (thus right-hand side of \refp{rel2} vanishes). Under these assumptions, 
Yang-Baxter equations \refp{YB1} and \refp{YB2} can be deduced from Jacobi identity of this algebra. 

Finally, remark that all these formulae are independent from the choice of $SL(2,R)/SO(2)$ coset 
model. The generalization to any $G/H$ coset is obvious : the tensors $c_{12}$ and $d_{12}$ have to be
 replaced by those of the new algebra. 

\section{Monodromy matrices}
\label{Mono}
Now that we have determined the Poisson brackets of the Lax connection in the previous section, we can go 
further and calculate those of the monodromy matrices. So we want to determine $\left\{\Psi_1(x,x_0),\Psi_2(y,y_0)\right\}$
, where $x > x_0$, $ y > y_0$ and the four points are distinct points. The way to achieve this calculation 
is as follows. We will use the Leibniz rule

\debut
\left\{\Psi_{ij}(x),\Psi_{kl}(y)\right\}&=&\int\!\!\int dz dz' \frac{\delta \Psi_{ij}(x)}{\delta A_{mn}(z)}
\left\{A_{mn}(z),A_{m'n'}(z')\right\}\frac{\delta \Psi_{kl}(y)}{\delta A_{m'n'}(z')}\non
\fin

The only difficulty is to find the functional derivative of $\Psi(x,x_0)$ 
with respect to $A(z)$. It can be done by solving the differential equations 
\debut
&\xder\delta\Psi(x,x_0)\;+\;A(x)\;\delta\Psi(x,x_0)\;+\;\delta A(x)\;\Psi(x,x_0)=0&\non\\
&\partial_{x_0}\delta\Psi(x,x_0)\;-\;\delta\Psi(x,x_0)\;A(x_0)\;-\;\Psi(x,x_0)\;\delta A(x_0)=0&\non
\fin
With the condition $\delta\Psi(x,x)=0$, the solution is given by
\debut
\delta\Psi(x,x_0)&=&\int^{+\infty}_{-\infty}\!\!dz\;\Theta(x-z)\;\Theta(z-x_0)\;\Psi(x,z)\;\delta A(z)\;\Psi(z,x_0)\non
\fin
where $\Theta(x)$ is the Heaviside function ($\Theta(x)=1$ for $x>0$, $\Theta(x)=0$ elsewhere).
The rest of the calculations is quite easy, consisting in putting together terms to form total derivatives. 
Thus, we obtain the following expression : 
\debut
\left\{\Psi_1(x,x_0),\Psi_2(y,y_0)\right\}&=&
        -\Theta(y,x,y_0)\;\rho^{-1}(x)\;\Psi_2(y,x) \;r^\epsilon_{12}\;
        \;\Psi_1(x,x_0)\Psi_2(x,y_0)\non\\
        &&-\Theta(x,y,x_0)\;\rho^{-1}(y)\;\Psi_1(x,y) \;s^\epsilon_{12}\;
        \Psi_1(y,x_0)\Psi_2(y,y_0)\non\\
        &&+\Theta(y,x_0,y_0)\;\rho^{-1}(x_0)\;\Psi_1(x,x_0) \Psi_2(y,x_0)
        \;r^\epsilon_{12}\; \Psi_2(x_0,y_0)\label{monobra}\\
        &&+\Theta(x,y_0,x_0)\;\rho^{-1}(y_0)\;\Psi_1(x,y_0) \Psi_2(y,y_0)
        \;s^\epsilon_{12}\; \Psi_1(y_0,x_0)\non\\
        &&-\frac{1}{8}\int^{+\infty}_{-\infty}\!\!dz\;
        \Theta(x-z)\Theta(z-x_0)\Theta(y-z)\Theta(z-y_0)\;\non\\
        &&\rho^{-2}(z)\;
        \Psi_1(x,z)\Psi_2(y,z)\;\left[U_{12},\Phi_1(z)-\Phi_2(z)\right]\;
        \Psi_1(z,x_0)\Psi_2(z,y_0)\non
\fin
with $\Theta(x,y,z)$ equal to 1 if $x>y>z$, and 0 for the other case. 

We can easily verify that these brackets are consistent. It is also a consequence of relations \refp{rel1},
\refp{YB1}, \refp{YB2} and \refp{rel2}. This last one gives the following relation for monodromy matrices 
\debut
\rho^{-1}(x)\;\left(r_{12}^\epsilon-r_{12}^{-\epsilon}\right)\;\Psi_1(x,y)\Psi_2(x,y)
&=&\rho^{-1}(y)\;\Psi_1(x,y)\Psi_2(x,y)\;\left(r_{12}^\epsilon-r_{12}^{-\epsilon}\right)\label{rel2mono}
\fin
We need also to evalute the brackets between $\Psi$ and $\rho$ that can be deduced form those of $A$ and $\rho$
\debut
\left\{A(x),\rho(y)\right\}&=&-\frac{1}{2}\;\delta(x-y)\;k\label{Arho}\\
\left\{\Psi(x,y),\rho(z)\right\}&=&\frac{1}{2}\;\Theta(x-z)\;k\;\Psi(x,y)\;-\;\frac{1}{2}\;\Theta(y-z)\;k\;\Psi(x,y)\label{Psirho}
\fin

What happens if we let the two ending points (or the two starting points) tend to the same value? In this case,
 we have to face to a well-known problem of non-ultralocal theories (see \cite{Maillet2} for example) that
 the brackets are ill-defined (in particular, the Jacobi identity is no longer valid). In the case of non-linear
 sigma models, it has been shown \cite{Forger} that no regularization of the Poisson brackets is coherent.
One way to go beyond this problem is to have additional informations about the boundaries conditions. For example,
 if we consider $\rho$ as the radial coordinate, we can choose a frame such that $\rho(x)$ tends to $\infty$ when
 $x$ goes to a given point $x_\infty$. If we take this naive limit in eq.\refp{monobra}, assuming that the $\Psi(x,x_\infty)$ terms
 have a good behavior compared to $\rho^{-1}(x)$, we obtain the following relation
\debut
\left\{\Psi_1(x,x_\infty),\Psi_2(y,x_\infty)\right\}&=&
        -\Theta(y-x)\;\rho^{-1}(x)\;\Psi_2(y,x_\infty)\Psi^{-1}_2(x,x_\infty) \;r^\epsilon_{12}\;
        \;\Psi_1(x,x_\infty)\Psi_2(x,x_\infty)\non\\
        &&-\Theta(x-y)\;\rho^{-1}(y)\;\Psi_1(x,x_\infty)\Psi_1^{-1}(y,x_\infty) \;s^\epsilon_{12}\;
        \Psi_1(y,x_\infty)\Psi_2(y,x_\infty)\non\\
        &&-\frac{1}{8}\int^{+\infty}_{-\infty}\!\!dz\;
        \Theta(x-z)\Theta(y-z)\;\rho^{-2}(z)\;\Psi_1(x,x_\infty)\Psi_2(y,x_\infty)\label{wavebra}\\
        &&\Psi_1^{-1}(z,x_\infty)\Psi_2^{-1}(z,x_\infty)\;\left[U_{12},\Phi_1(z)-\Phi_2(z)\right]\;
        \Psi_1(z,x_\infty)\Psi_2(z,x_\infty)\non
\fin
It can be shown that the above definition, these brackets are well-defined. We focus the readers attention on the 
consequences of this limit for relation \refp{rel2mono}. We obtain
\debut 
&\left(r_{12}^\epsilon-r_{12}^{-\epsilon}\right) \left(\rho^{-1}(x)\Psi_1(x,x_\infty)\Psi_2(x,x_\infty)\right)=0&\non
\fin
This boundaries condition implies that $\rho^{-1}\Psi_1\Psi_2$ has to be in the kernel of $r_{12}^\epsilon-r_{12}^{-\epsilon}$,
 which imposes strong constraints on $\Psi$.

Can we go further? If we keep in mind the equivalence of $\rho$ as the radial coordinate, it would seem interesting to
 consider the case $\rho=0$. It could be a way to define an algebra for spatially independent objects. Unfortunately,
 it leads to more difficult problems when trying to evaluate Poisson brackets of these objects. But as we shall see
 in the following section, such an approach is more successful when studying physical observables.

\section{Classical observables}
\label{Obs}

\subsection{General Framework}
\label{GenFram}

By definition, classical observables are functionals of the phase space variables that commute with the constraints.
 Before finding these observables, we need some preliminary calculations. First, we have to determine the commutators
 between the constraints and the connection :
\debut
\left\{{\cal{H}}(x),A_x(y)\right\}
&=&A_t(x)\xder\delta(x-y)-\left[A_x(x),A_t(x)\right]\delta(x-y)\non\\
	&&-\rho^{-1}(x)\left[\Phi(x),P_t(x)\right]\delta(x-y)\label{CHA}\\
\left\{{\cal{P}}(x),A_x(y)\right\}&=&A_x(x)\xder\delta(x-y)\label{CPA}\\
\left\{\Phi_1(x),A_{x2}(y)\right\}
        &=&\left[c_{12},A_{x2}(x)\right]\delta(x-y)
        +\xder\delta(x-y) c_{12}\label{CPhiA}
\fin
The equivalent relations for the wave function can be found by solving the differential equation associated to
 one of the commutators, obtained when derivating with respect to the spatial parameter of the wave function.
 Thus, we deduce the following identities (on the constraint surface):
\debut
\left\{{\cal{H}}(x),\Psi(y,z)\right\}&=&A_t(y)\Psi(y,z)\delta(x-y)-\Psi(y,z)A_t(z)\delta(x-z)\label{CHP}\\
\left\{{\cal{P}}(x),\Psi(y,z)\right\}&=&A_x(y)\Psi(y,z)\delta(x-y)-\Psi(y,z)A_x(z)\delta(x-z)\label{CPP}\\
\left\{\Phi_1(x),\Psi_2(y,z)\right\}&=&c_{12}\Psi_2(y,z)\delta(x-y)-\Psi_2(y,z)c_{12}\delta(x-z)\label{CPhiP}
\fin

Finding quantities that have vanishing brackets with the local $SO(2)$ constraint is quite obvious. If we
 consider $\zeta$ defined by $\partial_\mu\zeta+Q_\mu\zeta=0$, then it is straightforward to show that
 $\zeta^{-1}(x)\Psi(x,y)\zeta(y)$ commutes with $\Phi$. Difficulties arise when we try to find objects invariant
 under diffeomorphisms. If we keep in mind what was done in simpler cases, we should attempt to consider the
 monodromy matrix between the boundaries. We shall see that it can be achieved by using two particular values
 of $\rho$ and imposing physical boundary conditions to the solutions.

\subsection{Vacuum solution and level one representations}

We shall recall some formulae and results described in \cite{BerJul} that will be helpful when dealing with
 boundary conditions. In particular, we shall introduce the level one representations for the affine algebra, 

One of the most simple solution of the Einstein's equations is the vacuum solution which corresponds to the case
 where where all P and Q fields are null (notice there is a slightly change comparing to \cite{BerJul}, where
 $\hat{\sigma}$ is also zero). The associated Lax connection belongs to the Virasoro algebra
\debut
A_x&=&-\frac{1}{2}\rho^{-1}(\Pi_{\hat\sigma}-\xder \rho)E_+-
\frac{1}{2}\rho^{-1}(\Pi_{\hat\sigma}+\xder \rho)E_-+\Pi_{\rho}\frac{k}{2}\non\\
A_t&=&-\frac{1}{2}\rho^{-1}(\Pi_{\hat\sigma}-\xder \rho)E_++
\frac{1}{2}\rho^{-1}(\Pi_{\hat\sigma}+\xder \rho)E_--\xder\hat{\sigma}\frac{k}{2}\non
\fin
Solving \refp{lin} with this connection, we deduce the vacuum wave function $\Psi_V$ 
\debut
\Psi_V&=&e^{\frac{1}{2}\hat{\sigma}k}\left(\frac{\rho+\int{\Pi_{\hat\sigma}}+b}
{2\rho}\right)^{E+}\left(\frac{\rho+\int{\Pi_{\hat\sigma}}+b}{b'}\right)^{E-}\label{PsiV}\\
&=&e^{\frac{1}{2}\hat{\sigma}k}\left(\frac{\rho-\int{\Pi_{\hat\sigma}}+c}
{2\rho}\right)^{-E-}\left(\frac{\rho-\int{\Pi_{\hat\sigma}}+c}{c'}\right)^{-E+}\non
\fin

We shall introduce the level one representations as a tool to specify physical
 observables. First we introduce the two dimensional representation of $sl(2,R)$ involving Pauli matrices
\debut
T^x=\frac{1}{\sqrt{2}}\left(\begin{array}{cc} 0&1\\1&0\end{array}\right)&
T^y=\frac{1}{\sqrt{2}}\left(\begin{array}{cc} 0&i\\-i&0\end{array}\right)&
T^z=\frac{1}{\sqrt{2}}\left(\begin{array}{cc} 1&0\\0&-1\end{array}\right)\non
\fin
Notice that $T^y$ is the $SO(2)$ generator.

Let denote $Z(\mu)$ the field
\debut
Z(\mu)=\sum_{n\;\;{\rm{odd}}}p_{-n}\frac{\mu^n}{n}&{\rm{with}}&[p_n,p_m]=n\delta_{n+m,0}\non
\fin
and the vertex operator $W_2(\mu)=:e^{-2iZ(\mu)}:$. We denote $|0>$ the vacuum of the Fock space
 generated by the operators $p_n$. The level one representations with highest weight $\Lambda_\pm$ are defined by
\debut
i\mu\frac{dZ(\mu)}{d\mu}&=&\sum_{n\;\;{\rm{odd}}}(T^z \otimes \lambda^n)\mu^{-n}\non\\
\pm i\;\;W_2(\mu)&=&2\sum_{n\;\;{\rm{even}}}(T^y \otimes \lambda^n)\mu^{-n}-2\sum_{n\;\;{\rm{odd}}}(T^x \otimes \lambda^n)\mu^{-n}\non
\fin
The heighest weight vectors $|\Lambda_\pm>$ are identified with $|0>$. The Virasoro generators are represented by
\debut
L_n&=&-\frac{1}{8\pi i}\oint_{\cal{C}}{d\mu\;\mu^{2n+1}:(i\partial_\mu Z)^2:}\;\;+\;\;\frac{1}{16}\delta_{n,0}\non
\fin
where $\cal{C}$ is a contour around zero.

In the next subsection, we shall need to conjugate elements of the algebra by the vacuum wave function.
 We gather here all needed formulae. The case of the Virasoro algebra elements $E_+$ and $E_-$ is quite obvious
\debut
D^{E_+}\;E_-\;D^{-E_+}=D\;E_-+(D-1)E_+&{\rm{and}}&B^{-E_-}\;E_+\;B^{E_-}=B\;E_++(B-1)E_-\non
\fin
Operators $\mu^{-1}W_2(\mu)$ and $\partial_\mu Z$ are primary fields of weight 1. Under a diffeomorphism
 $\mu\rightarrow F(\mu)$, they transform as
\debut
\mu^{-1}W_2(\mu)&\rightarrow&\frac{\mu}{2F^2} \left(\partial_\mu F^2\right) W_2(F(\mu))\non\\
\partial_\mu Z&\rightarrow&\frac{\mu^2}{2F^2} \left(\partial_\mu F^2\right) (\partial_\mu Z)(F(\mu))\non
\fin
It can be proved that conjugations $AD^{-E_+}\mu^{-1}W_2(\mu)D^{E+}$ and $B^{E_-}\mu^{-1}W_2(\mu)B^{-E_-}$,
 are associated to the following diffeomorphisms
\debut
F^2_+(\mu)&=&\frac{\mu^2}{\mu^2+(1-\mu^2)D}\non\\
F^2_-(\mu)&=&1+(\mu^2-1)B\non
\fin

\subsection{Boundaries conditions and physical observables}

Let give a brief sketch of the strategy we shall use. Following the idea proposed in \cite{KoroSamt1}, we
 shall consider $\Psi$ between the points $x_0$ where $\rho=0$, and $x_\infty$ where $\rho\rightarrow\infty$
. We shall need to define our phase space by specifying the behavior of the other fields in these limits. As
 we will see, it is not possible to obtain physical observables by only imposing boundaries conditions on $\Psi$.
 We shall construct physical observables in the form $M_0^{-1}(x_0)\Psi(x_0,x_\infty)M_\infty(x_\infty)$ in such
 a way the contribution of $M_0$ and $M_\infty$ to the Poisson brackets with the constraints eliminates the unwanted terms.

First we shall look at the case where $\rho\rightarrow\infty$, with the picture that in this limit $\rho$ becomes
 equivalent to the usual radial coordinate (assuming it happens when $x\rightarrow x_\infty$). It seems
 physically reasonable to restrict our phase space to solutions which tend asymptotically to the flat space
 solution when  $\rho\rightarrow\infty$. This solution corresponds to the Minkowski metric element expressed in
 cylindrical coordinates $ds^2=-dt^2+d\rho^2+\rho^2d\theta^2+dz^2$. We shall take 
\debut
\Pi_\rho=\Pi_{\hat\sigma}=0,&\hat{\sigma}=-\frac{1}{4}\ln\rho,&P_x=\rho^{-1}\sqrt{2}T^z\non
\fin
and all other components of $P$ and $Q$ equal to zero. The components of the Lax connection are
\debut
(A_f)_x&=&\rho^{-1}\left(\frac{1}{2}(E_+-E_-)+\frac{\sqrt{2}}{2}T^z(\lambda+\lambda^{-1})\right)\label{Afx}\\
(A_f)_t&=&\rho^{-1}\left(\frac{1}{2}(E_++E_-)+\frac{\sqrt{2}}{2}T^z(\lambda-\lambda^{-1})+\frac{k}{8}\right)\label{Aft}
\fin
The wave function which is solution of \refp{lin} with the above connection can be written as
\debut
\Psi_f&=&h_-\left(\frac{2\rho}{\rho+1}\right)^{E_-}\left(\frac{1}{2}(\rho+1)\right)^{-E_+}\rho^{-\frac{k}{2}}h_-^{-1}\non\\
      &=&h_+\left(\frac{2\rho}{\rho+1}\right)^{-E_+}\left(\frac{1}{2}(\rho+1)\right)^{E_-}\rho^{\frac{k}{2}}h_+^{-1}\non
\fin
with $h_\pm=\exp\left(\sqrt{2}T^z\ln\left(\frac{1+\lambda^{\pm 1}}{1-\lambda^{\pm 1}}\right)\right)$. If we look at the
 Lax connection, we see that it is proportional to $\rho^{-1}$. Thus, if we impose that the wave function $\Psi(x)$ tends
 asymptotically toward $\Psi_f(x)$ when $x\rightarrow x_\infty$, equations \refp{CHP} and \refp{CPP} become
\debut
\left\{{\cal{H}}(x),\Psi(y,x_\infty)\right\}&=&A_t(y)\Psi(y,x_\infty)\delta(x-y)\label{CHinf}\\
\left\{{\cal{P}}(x),\Psi(y,x_\infty)\right\}&=&A_x(y)\Psi(y,x_\infty)\delta(x-y)\label{CPinf}
\fin

Now let us consider the more tricky case of $\rho=0$. We shall suppose there is a point $x_0$ such that 
\debut
\rho(x_0)=0&{\rm{and}}&\Pi_{\hat{\sigma}}(x_0)=0\non
\fin
In the picture of cylindrical symmetry, it means that close to the symmetry axis, $\rho$ can be identified with
 the usual radial coordinate. We assume $P$ behaves like $\rho^{-1}$ when $x\rightarrow x_0$. Actually, if we
 admit these quantities have a $\rho$ power law behavior in this limit (which seems physically reasonable), then,
 using equations of motion (\ref{motion1}, \ref{motion2}, \ref{motion3}), we can show that it is the only solution.
 But the monodromy matrix $\Psi(x_0,x_\infty)$ is not an observable because the Lax connection diverges when $\rho\rightarrow 0$.
 To avoid this problem, we apply the technique explained at the beginning of this subsection: we multiply
 $\Psi(x,x_\infty)$ by $M_0^{-1}(x)$ with $M_0$ equal to the vacuum wave function $\Psi_V$ whose form in the
 limit $x\rightarrow x_0$ becomes
\debut
\Psi_V&\sim e^{\frac{1}{2}\hat{\sigma}k}\left(\frac{b}{2\rho}\right)^{E+}\left(b'\right)^{-E-}&
\sim e^{\frac{1}{2}\hat{\sigma}k}\left(\frac{c}{2\rho}\right)^{-E-}\left(c'\right)^{E+}\label{PsiV0}
\fin
(relabeling the prime constants to simplify formulae).

The linear equations satisfied by $\Psi_V^{-1}(x)\Psi(x,x_\infty)$ are of type
\debut
\partial\left(\Psi_V^{-1}(x)\Psi(x,x_\infty)\right)&=&\left(\Psi_V(x)^{-1}\tilde{A}(x)\Psi_V(x)\right)\Psi_V(x)^{-1}\Psi(x,x_\infty)\non
\fin
where $\tilde{A}$ is the  Kac-Moody algebra part of the connection (the contribution of $\Psi_V$ has canceled Virasoro
 and central extension parts). Poisson brackets of $\Psi_V^{-1}(x)\Psi(x,x_\infty)$ with $\cal{H}$ and $\cal{P}$ are
 obtained by substituting $\Psi_V^{-1}\tilde{A}\Psi_V$ to $A$ in formulae \refp{CHinf} and \refp{CPinf}. To evaluate
 $\Psi_V^{-1}\tilde{A}\Psi_V$, we use the level one representation described in the previous subsection and the second
 form of \refp{PsiV0}. The diffeomorphism associated to this conjugation is 
$F^2(\mu)=\frac{2\rho +(\mu^2-1)c}{2\rho +(\mu^2-1)c(1-c')}$. For example, for one of the components of $A$, we have
\debut
\Psi_V^{-1}\;(P_t)_xT^x\otimes\lambda\;\Psi_V&=&\mp\frac{(P_t)_x}{4\pi}\Psi_V^{-1}
\;\left(\oint_{\cal{C}}{d\mu\;\mu W_2(\mu)}\right)\;\Psi_V\non\\
&=&\mp\frac{(P_t)_x}{4\pi}\left(\rho \frac{c'}{c(c'-1)^2}W_2(\frac{1}{1-c'})
\oint_{\cal{C}}{d\mu\;\frac{\mu^3}{(\mu^2-1)^2}}+{\cal{O}}(\rho^2)\right)\non\\
&=&(P_t)_x\left(0+{\cal{O}}(\rho^2)\right)={\cal{O}}(\rho)\;\;{\rm{as}}\;\;\rho\rightarrow 0\non
\fin
The last equation follows from the fact that $P$ is proportional to $\rho^{-1}$ as $x\rightarrow x_0$ and
 that the contour $\cal{C}$ around $0$ can be chosen as small as we want, such that there is no pole contribution
 to the integral. Doing these calculations in the level one representation provides a way to give meaning to
 quantities like $W_2(\frac{1}{1-c'})$, which would be ambiguous in the abstract Kac-Moody algebra.

Identical results can be obtained for the other components. Thus $\Psi_V^{-1}(x_0)\tilde{A}(x_0)\Psi_V(x_0)$ is
 equal to zero, whereas $A(x)$ was divergent when $x\rightarrow x_0$, and $\Psi_V^{-1}(x_0)\Psi(x_0,x_\infty)$
 has null Poisson brackets with the generators of diffeomorphisms.

Defining $SO(2)$ gauge-invariant object from $\Psi_V^{-1}(x_0)\Psi(x_0,x_\infty)$ is quite straightforward
. Using the technique presented in subsection \ref{GenFram} and the fact that $so(2)$ elements commute with
 $\Psi_V$, we can show that the quantity 
\debut
\tilde{\Psi}(x_0,x_\infty)&=&\zeta^{-1}(x_0)\Psi_V^{-1}(x_0)\Psi(x_0,x_\infty)\zeta(x_\infty)\label{Psiobs}
\fin
still has vanishing brackets with ${\cal{H}}$ and ${\cal{P}}$, and is moreover $SO(2)$ gauge-invariant. It
 proves that $\tilde{\Psi}(x_0,x_\infty)$ is a physical observable. Note that these are operators (ie infinite
 dimensional matrices) acting on the level one representation of $sl(2,R)$. They thus provide an infinite set
 of physical observables.

\bigskip

{\bf To summarize:} We have supposed the two following boundary conditions: 1) when $\rho$ goes to infinity,
 the wave function tends asymptotically to the flat space wave function; 2) we can find a point $x_0$ where $\rho$
 and $\Pi_{\hat{\sigma}}$ are null. Notice that these conditions are fulfilled in concrete examples like cylindrical
 gravitational waves (see e.g. \cite{Kuchar}). Using these hypotheses and the level one representations, we have
 shown that $\zeta^{-1}(x_0)\Psi_V^{-1}(x_0)\Psi(x_0,x_\infty)\zeta(x_\infty)$ generates an infinite set of classical
 physical observables. It remains to decipher the Poisson bracket algebra they generate which should be closer to those
 of the Toda's theories.

\section{Appendix A:}
\label{AppA}

The aim of this appendix is to give more details about the Poisson brackets of the connection. First, let us write
 developed formulae for the $r-$ and $s-$matrices :
\begin{itemize}
\item plus convention $\left(|\lambda_1|<|\lambda_2|\right)$:
\debut
r^+_{12}&=&\frac{1}{2}(1-\lambda_1^{2})(1-\lambda_2^{-2})
        \sum_{n \geq 0}\left(\frac{\lambda_1}{\lambda_2}\right)^{2n} c_{12}
        -\frac{1}{2}(\lambda_2-\lambda_2^{-1})^2
        \sum_{n \geq 0}\left(\frac{\lambda_1}{\lambda_2}\right)^{2n+1} d_{12}\non\\
        &&-\frac{1}{2}E_+ \otimes k-\frac{1}{4}k \otimes 
        \left(E_++E_-\right) \non\\
s^+_{12}&=&\frac{1}{2}(1-\lambda_1^{2})(1-\lambda_2^{-2})
        \sum_{n \geq 0}\left(\frac{\lambda_1}{\lambda_2}\right)^{2n} c_{12}
        -\frac{1}{2}(\lambda_1-\lambda_1^{-1})^2
        \sum_{n \geq 0}\left(\frac{\lambda_1}{\lambda_2}\right)^{2n+1} d_{12}\non\\
        &&-\frac{1}{2}k \otimes E_--\frac{1}{4} 
        \left(E_++E_-\right) \otimes k \non
\fin
\item minus convention $\left(|\lambda_1|>|\lambda_2|\right)$:
\debut
r^-_{12}&=&-\frac{1}{2}(1-\lambda_1^{-2})(1-\lambda_2^{2})
        \sum_{n \geq 0}\left(\frac{\lambda_2}{\lambda_1}\right)^{2n} c_{12}
        +\frac{1}{2}(\lambda_2-\lambda_2^{-1})^2
        \sum_{n \geq 0}\left(\frac{\lambda_2}{\lambda_1}\right)^{2n+1} d_{12}\non\\
        &&+\frac{1}{2}E_- \otimes k + \frac{1}{4}k \otimes 
        \left(E_++E_-\right) \non\\
s^-_{12}&=&-\frac{1}{2}(1-\lambda_1^{-2})(1-\lambda_2^{2})
        \sum_{n \geq 0}\left(\frac{\lambda_2}{\lambda_1}\right)^{2n} c_{12}
        +\frac{1}{2}(\lambda_1-\lambda_1^{-1})^2
        \sum_{n \geq 0}\left(\frac{\lambda_2}{\lambda_1}\right)^{2n+1} d_{12}\non\\
        &&+\frac{1}{2}k \otimes E_++\frac{1}{4} 
        \left(E_++E_-\right) \otimes k \non
\fin
\end{itemize}

Notice that the differences $r^+_{12}-r^-_{12}$ and $s^+_{12}-s^-_{12}$ are equal. But in contrary
 to the usual cases, they are not proportional to the Casimir tensor (see equation (\ref{rel2})).

In order to demystify this formulae, we will sketch the way we have obtained them. First, we consider only
 the loop part of the algebra and the ultralocal contribution. The more general expression for the two
 $r-$matrices we can take, is of type $f(\lambda_1,\lambda_2)c_{12}+g(\lambda_1,\lambda_2)d_{12}$. Comparing
 with the raw formula of the Poisson brackets, we deduce the loop part of (\ref{r+-}) and (\ref{s+-})
\debut
r_{12}=f(\lambda_1,\lambda_2)c_{12}+g(\lambda_1,\lambda_2)d_{12}&{\rm{and}}&s_{12}
=f(\lambda_1,\lambda_2)c_{12}-g(\lambda_2,\lambda_1)d_{12}
\fin 
with 
\debut
f(\lambda_1,\lambda_2)=\frac{1}{2}\frac{(1-\lambda_1^2)(1-\lambda_2^2)}{\lambda_1^2-\lambda_2^2}&{\rm{and}}&
g(\lambda_1,\lambda_2)=\frac{1}{2}\frac{\lambda_1\lambda_2^{-1}(1-\lambda_2^2)^2}{\lambda_1^2-\lambda_2^2}\non
\fin
These rational functions verify some non-trivial algebraic relations that are helpful when dealing with Jacobi identity
\debut
g(\lambda_1,\lambda_3)g(\lambda_3,\lambda_2)+f(\lambda_2,\lambda_3)g(\lambda_1,\lambda_2)
+f(\lambda_3,\lambda_1)g(\lambda_1,\lambda_2)&=&0\non\\
g(\lambda_1,\lambda_2)g(\lambda_3,\lambda_2)-f(\lambda_2,\lambda_3)g(\lambda_1,\lambda_3)
-f(\lambda_1,\lambda_2)g(\lambda_1,\lambda_3)&=&0\non\\
g(\lambda_1,\lambda_2)g(\lambda_1,\lambda_3)-f(\lambda_1,\lambda_3)g(\lambda_3,\lambda_2)
-f(\lambda_1,\lambda_2)g(\lambda_2,\lambda_3)&=&(\lambda_2-\lambda_2^{-1})(\lambda_3-\lambda_3^{-1})\non\\
\fin

Now we have to add the central extension (which impose to choose a convention for the previous equations).
 We can easily see that we need also to introduce the Virasoro algebra. The only possible and non-trivial
 terms are $E_\pm \otimes k$ and $k \otimes E_\pm$. The goal we want to reach is to include all terms whose
 variables are different from the dilaton and its spatial derivative, into the two commutators
 $\left[r^\epsilon_{12},A_1(x)\right] + \left[s^\epsilon_{12},A_2(x)\right]$. The first miracle is that it
 can be achieved. Thus, we fix one part of the $r-$matrices on the Virasoro algebra (remember that $k$ is in
 the center of the algebra, so $\left[k \otimes E_\pm,A_1(x)\right]=0$ and so on). The last step consists
 in fixing the rest of the $r-$matrices in such a way that we obtain a compact form for (\ref{fpb}). And here
 we have a second miracle: it can be done for the ultralocal and the non-ultralocal part simultaneously.\\
We see this formula works due to non-trivial cancelations. In one sense, it's a proof of the validity of our
 calculation and a hint of deeper algebraic structure of the problem.

\section{Appendix B:}
\label{AppB}

Here we will give some hints for the proof of the Jacobi identity. First of all, let recall the expression of eq.\refp{Jac} :
\debut
\left\{A_1(x),\left\{A_2(y),A_3(z)\right\}\right\}+{\rm{perm.}}
        &=&\rho^{-2}(x)\delta(x-y)\delta(y-z)\left(
        \left[A_1(x),A_{123}\right]+{\rm{perm.}}\right)\non\\
        &+&\left(\xder\rho^{-2}\right)\delta(x-y)\delta(y-z)
        \left(B_{123}+{\rm{perm.}}\right)\non \\
        &+&\rho^{-3}(x)\delta(x-y)\delta(y-z)\left(
        \left[\Phi_1(x),C_{123}\right]+{\rm{perm.}}\right)\non\\
        &+&\rho^{-2}(x)\delta(y-z)\xder\delta(x-y)D_{123}+{\rm{perm.}}
        \non    
\fin
with the following values for the coefficients
\debut
A_{123}&=&[r_{12}^{\epsilon_1},r_{23}^{\epsilon_2}]
        +[s_{23}^{\epsilon_2},s_{31}^{\epsilon_3}]
        +[s_{31}^{\epsilon_3},r_{12}^{\epsilon_1}]
        -\frac{1}{2}k_2s^{\epsilon_3}_{31}-\frac{1}{2}k_3r^{\epsilon_1}_{12}
        -\frac{1}{4}\left[U_{23},c_{12}\right]\non\\
B_{123}&=&[s_{23}^{\epsilon_2},s_{31}^{\epsilon_3}]
        -[r_{23}^{\epsilon_2},r_{12}^{\epsilon_1}]
        +\frac{1}{2}[r_{23}^{\epsilon_2},s_{12}^{\epsilon_1}]
        -\frac{1}{2}[s_{23}^{\epsilon_2},r_{31}^{\epsilon_3}]
        -\frac{1}{4}\left[U_{23},c_{12}\right]\non\\
C_{123}&=&\frac{1}{4}[r_{23}^{\epsilon_2},U_{12}]
        +\frac{1}{4}[s_{23}^{\epsilon_2},U_{31}]
        +\frac{1}{8}k_3U_{12}-\frac{1}{8}k_2U_{13}\non\\
D_{123}&=&[s_{23}^{\epsilon_2},s_{31}^{\epsilon_3}]
        -[r_{23}^{\epsilon_2},r_{12}^{\epsilon_1}]
        +[r_{23}^{\epsilon_2},s_{12}^{\epsilon_1}]
        -[s_{23}^{\epsilon_2},r_{31}^{\epsilon_3}]
        -\frac{1}{4}\left[U_{23},c_{12}\right]\non\\
        &&+\frac{1}{4}k_3\left(s_{12}^{\epsilon_1}-r_{12}^{\epsilon_1}\right)
        -\frac{1}{4}k_2\left(s_{31}^{\epsilon_3}-r_{31}^{\epsilon_3}\right)
        +\frac{1}{4}k_1\left(s_{23}^{\epsilon_2}+r_{23}^{\epsilon_2}\right)\non
\fin
To obtain this expression, we need Poisson brackets of connection with the constraint $\Phi$
 and the dilaton (see eq.\refp{CPhiA} and \refp{Arho}).

Showing Jacobi identity with \refp{Jac} is not obvious, because the three $\delta \partial \delta$
 distributions and $\delta \delta$ are not linearly independent. So, we have to express one of
 $\delta \partial \delta$ with respect to the other distributions. It leads to the following formula :
\debut
\left\{A_1(x),\left\{A_2(y),A_3(z)\right\}\right\}+{\rm{perm.}}
        &=&\rho^{-2}(x)\delta(x-y)\delta(y-z)\left(
        \left[A_1(x),A_{123}\right]+{\rm{perm.}}\right)\non\\
        &+&\left(\xder\rho^{-2}\right)\delta(x-y)\delta(y-z)
        \left(B_{123}+B_{231}+B_{312}-2D_{123}\right)\non \\
        &+&\rho^{-3}(x)\delta(x-y)\delta(y-z)\left(
        \left[\Phi_1(x),C_{123}\right]+{\rm{perm.}}\right)\non\\
        &+&\rho^{-2}(y)\delta(z-x)\partial_y\delta(y-z)\left(D_{123}-D_{123}\right)\non\\
        &+&\rho^{-2}(z)\delta(x-y)\partial_z\delta(y-z)\left(D_{312}-D_{123}\right)
        \non    
\fin
It can be easily shown that we have the relations
\debut
&B_{123}+B_{231}+B_{312}-2D_{123}=A_{231}+A_{312}-A_{123}&\non\\
&D_{123}-D_{231}=A_{123}-A_{231}&\non\\
&D_{123}-D_{312}=A_{123}-A_{312}&\non
\fin
Thus, our problem is entirely expressed in terms of $A_{123}$ and $C_{123}$. As announced previously,
 these coefficients are invariant under cyclic permutations and equal to zero :
\debut
A_{123}=A_{231}=A_{312}=0&{\rm{and}}&C_{123}=C_{231}=C_{312}=0\non
\fin
If the calculations for the $C$ coefficients are rather easy, those for the $A$ ones are more tedious.
 In particular, when dealing with terms of type $k \otimes d$, the choice of the convention has to be
 coherent. Another point is the fundamental Poisson brackets are not to be taken on the constraint surface.
 Else the $[U,c]$ term disappears, and the equation for $A$ on the loop part of the algebra is no longer verified.


\newpage

\begin{thebibliography}{}

\bibitem{Geroch} R. Geroch, J. Math. Phys. 13 (1972) 394.

\bibitem{BelZak} V.A. Belinskii and V.E. Zakharov, Sov. Phys. 48 (1978) 985.

\bibitem{Maison} D. Maison, Phys. Rev. Lett. 41 (1978) 521.

\bibitem{NicoKoroSamt1} H. Nicolai, D. Korotkin and H. Samtleben, hep-th/9612065.

\bibitem{KoroSamt1} D. Korotkin and H. Samtleben,  Nucl. Phys. B527 (1998) 657.

\bibitem{JulNico} B. Julia and H. Nicolai, Nucl. Phys. B482 (1996) 431.

\bibitem{BerJul} D. Bernard and B. Julia, Nucl. Phys. B547 (1999) 427.

\bibitem{BerReg} D. Bernard and N. Regnault, Comm. Math. Phys. 210 (2000) 177

\bibitem{Maillet1} J.M. Maillet, Phys. Lett. B167 (1986) 401.

\bibitem{Maillet2} J.M. Maillet, Nucl. Phys. B269 (1986) 54.

\bibitem{Forger} M. Forger, M. Bordermann, J. Laartz, U. Schaper, Comm. Math. Phys. 152 (1993) 167

\bibitem{Kuchar} K. Kuchar, Phys. Rev. D4 (1971) 995

\end{thebibliography}
\end{document}